\documentclass[12pt]{article}
\usepackage{amsmath,amssymb,graphicx} 
\usepackage{epsf}
\usepackage{epsfig}
\usepackage{pstricks}
\usepackage{cite}

\newcommand{\be}{\begin{equation}}
\newcommand{\ee}{\end{equation}}

\newcommand{\beq}{\begin{eqnarray}}
\newcommand{\eeq}{\end{eqnarray}}

\newcommand{\centeron}[2]{{\setbox0=\hbox{#1}\setbox1=\hbox{#2}\ifdim
\wd1>\wd0\kern.5\wd1\kern-.5\wd0\fi
\copy0

\kern-.5\wd0\kern-.5\wd1\copy1\ifdim\wd0>\wd1
                                       \kern.5\wd0\kern-.5\wd1\fi}}
\newcommand{\ltap}{\>\centeron{\raise.35ex\hbox{$<$}}
                               {\lower.65ex\hbox{$\sim$}}\>}
\newcommand{\gtap}{\>\centeron{\raise.35ex\hbox{$>$}}
                               {\lower.65ex\hbox{$\sim$}}\>}

\newcommand\ZZ{\hbox{\zfont Z\kern-.4emZ}}
\font\zfont = cmss10 

\begin{document}
\begin{titlepage}
\begin{flushright}
\end{flushright}

\vskip.5cm
\begin{center}
{\huge \bf 
DAMA and WIMP dark matter
}

\vskip.1cm
\end{center}
\vskip0.2cm

\begin{center}
{\bf
Frank J. Petriello, 
Kathryn M. Zurek}
\end{center}
\vskip 8pt

\begin{center}
{\it Physics Department, University of Wisconsin, Madison, WI 53706} \\

\vspace*{0.1cm}
\vspace*{0.3cm}
{\tt frankjp@physics.wisc.edu, kzurek@wisc.edu}
\end{center}

\vglue 0.3truecm

\begin{abstract}
\vskip 3pt
\noindent

We study whether spin-independent scattering of weakly-interacting massive particles (WIMPs) with nuclei can account for the annual modulation signal reported 
by DAMA.  We consider both elastic and inelastic scattering processes.  We find that there is a region of WIMP parameter space which can simultaneously accommodate DAMA and the null results of CDMS, CRESST, and XENON.  This region corresponds to an ordinary, elastically-scattering WIMP with a standard Maxwell-Boltzmann distribution, a mass $3 \mbox{ GeV} \lesssim m_{DM} \lesssim 8 \mbox{ GeV}$, and a spin-independent cross section with nucleons $3 \times 10^{-41} \mbox{ cm}^2 \lesssim  \sigma_p^{SI} \lesssim 5 \times 10^{-39} \mbox{ cm}^2$.  This new region of parameter space depends crucially on the recently discovered effect of channeling on the energy threshold for WIMP detection in the DAMA experiment; without the inclusion of this effect, the DAMA allowed region is essentially closed by null experiments.  Such low-mass WIMPs arise in many theories of Beyond the Standard Model physics, from minimal extensions of the MSSM to solutions of the baryon-dark matter coincidence problem.   
We find that inelastic scattering channels do not open 
up a significant parameter region consistent with all experimental results.  Future experiments with low energy thresholds for detecting nuclear recoils, such as CDMSII-Si and those utilizing ultra-low energy germanium detectors, will be able to probe the DAMA region of parameter space.

\end{abstract}

\end{titlepage}

\newpage


\section{Introduction}
\label{sec:intro}

Recently, the DAMA collaboration has provided further evidence for the observation of an annual modulation in the rates of nuclear recoil in their experiment \cite{Bernabei:2008yi}.  Such a signal arises naturally from postulating Weakly Interacting Massive Particles (WIMPs) in the galactic halo that scatter from target nuclei in detectors.  The annual modulation of the interaction rate comes from the variation in the relative velocity of the earth with respect to the galactic dark matter halo as the earth orbits the sun.  This changes the flux of dark matter particles and the size of their interaction cross-sections, with expected extrema occurring at June 2 and December 2.  The DAMA experiment 
observes a maximum at low nuclear recoil energies on May 24, plus or minus 8 days, and they have accumulated enough data to put the significance of the observed modulation at approximately $8\sigma$.   Both the phase and amplitude of the signal are highly suggestive of WIMP interactions.  The collaboration has not been able to 
identify other systematic effects capable of producing this signal, and have claimed that the annual modulation is a discovery of dark matter.  This claim has been 
controversial, partly because a number of other experiments seem to be in direct contradiction.  In particular, the original DAMA allowed region with WIMP mass $30 \mbox{ GeV} \lesssim m_{DM} \lesssim 200 \mbox{ GeV}$ and dark matter-nucleon interaction cross-section $\sigma_p \simeq 10^{-41}-10^{-42} \mbox{ cm}^2$ had been quite conclusively ruled out by the CDMS~\cite{CDMSI,CDMSII,CDMSIISi} and XENON~\cite{XENON} experiments for the case of an elastically scattering WIMP.

Methods of reconciling the DAMA signal with the results of other experiments have been proposed in the past.  Inelastic scattering processes, $\chi_1 N \to 
\chi_2 N$, where $\chi_1$ is the dark matter particle and $\chi_2$ is another new state with mass splitting $\delta$ between the states, have been proposed in the context of supersymmetric models~\cite{Hall:1997ah,Smith:2001hy,TuckerSmith:2004jv}.  Inelastic scattering of MeV dark matter particles to lighter states was investigated as a possible solution in 
Ref.~\cite{Bernabei:2008mv}.  Mirror states from a hidden-sector copy of the Standard Model have been proposed as a candidate consistent with all experimental 
constraints~\cite{Foot:2005ic,Foot:2008nw}, as have various models with heavy composite states~\cite{Khlopov:2008ki}.  A model-independent study of spin-independent elastic scattering noted that scattering from the sodium component 
of the NaI DAMA scintillators allowed a small window of dark matter masses in the $5-9\,{\rm GeV}$ to be consistent with current experimental constraints~\cite{Gelmini,Gondolo:2005hh}.  This study noted that since sodium nuclei are lighter than germanium nuclei, the threshold for scattering off sodium could be lower than that for 
germanium for light dark matter states.  It was also shown previously that spin-dependent scattering may open up additional parameter space consistent with DAMA and other experiments~\cite{kami}.

This model-independent study of elastic scattering in Ref.~\cite{Gondolo:2005hh} is no longer applicable, as several new results have appeared recently.  New experimental 
constraints and improved understanding of scattering processes in the DAMA apparatus have drastically altered both the excluded parameter space and the physics underlying the DAMA modulation signal.   We perform a model-independent study of both elastic and inelastic scattering mechanisms accounting for all 
recent experimental measurements.  We find that completely ordinary, spin-independent elastically scattering WIMPs with masses in the range $3-8\,{\rm GeV}$ and scattering 
cross sections in the range $3\times 10^{-41}\,{\rm cm^2}$ to $5\times 10^{-39}\,{\rm cm^2}$ are consistent 
with all experimental constraints.  No additional dark matter stream is needed; a simple Maxwell-Boltzmann distribution allows this parameter space.  Inelastic scattering no longer opens up a significant region of additional allowed parameter space.  We summarize here the important features and conclusions of our analysis.
 
 \begin{itemize}

\item We include the effect of channeling in the NaI crystal scintillators of DAMA, an effect recently noted in Ref.~\cite{Drobyshevski:2007zj} and studied by the DAMA collaboration~\cite{Bernabei:2007hw}.  Channeling occurs in crystalline detectors where the only signal measured is the light output, and when recoiling nuclei interact only electromagnetically with the detector material because of either their direction of motion or incident energy.  The effect of channeling is to remove the quenching factor usually required to convert between nuclear recoil energy and electron-equivalent energy, and in the context of dark matter searches it effectively lowers 
the energy threshold for detection of nuclear recoils of DAMA below that of CDMS and XENON.  This effect is crucial in reconciling elastically scattered WIMPs consistent with all experimental constraints.  In particular, the lower threshold of DAMA means that it can detect lighter dark matter particles than the higher threshold experiments like CDMS and XENON.  This effect opens the region of light WIMP parameter space for DAMA.  We note that the presence of channeling in the energy regime studied by DAMA has not been conclusively established, although it has been 
observed in NaI crystals at higher energies~\cite{channel}.

\item We include constraints from CDMS-SUF, CDMS-II, CRESST-I scattering from sapphire targets~\cite{CRESST}, 
and XENON.  Inclusion of experimental results from multiple target nuclei is necessary to correctly elucidate the allowed parameter region.   In the elastic scattering case we also include the recent results from the CoGeNT collaboration~\cite{cogent}.

\item We study spin-independent elastic scattering, and inelastic scattering with either positive or negative mass splitting $\delta$ between the incident and scattered dark matter particle.  Inelastic scattering of either sign opens up 
only a very small window of parameter space; roughly, inelastic scattering to heavy states is ruled out by XENON and the germanium data from CDMS-II, while 
scattering to lighter states is ruled out by CRESST
and the silicon data from CDMS.  The preferred parameter space is for light mass dark matter with elastic scattering.  Future results from ultra-low noise germanium detectors~\cite{Barbeau:2007qi}~\footnote{We thank J. Collar for correspondence regarding the ability of these experiments to probe this region.}, and the lower threshold silicon data from CDMS will be vital in exploring this region.  The low threshold germanium experiment TEXONO~\cite{TEXONO} may also be able to probe this region, though its sensitivity has recently been called into question~\cite{Avignone:2008xc} (inclusion of current TEXONO constraints does not change our results).  

\end{itemize}

There are many possible models which could give rise to such a comparatively light WIMP.  In extensions of the Minimal Supersymmetric Model, for example, GeV mass WIMPs with the right relic abundance arise~\cite{Gunion,Barger}.  It has been shown that hidden sectors in the context of supersymmetric models give rise naturally to WIMPs with GeV or even lighter masses, as observed in Refs.\cite{Hooper,Feng}.  It was shown explicitly that the models of this sort can account for the DAMA signal~\cite{Feng:2008dz}.  Supersymmetric models with non-unified gaugino masses at the 
grand unified scale give rise to light neutralino dark matter candidates~\cite{Bottino:2002ry}; the importance of the channeling effect for these models was noted in 
Ref.~\cite{Bottino}, as was the effect of having the light sodium component of the DAMA target.  Lastly, solutions to the baryon-dark matter asymmetry problem also predict a WIMP with a mass in the range, $m_{dm} \approx \Omega_{dm}/\Omega_b m_p \approx 5 m_p$~\cite{Kaplan,Farrar,Kitano,fw}.  It is clear that the light dark matter paradigm suggested by the direct detection experiments raises numerous theoretical questions 
and has phenomenological impact on a broad array of experiments.  We leave the potential implications of light WIMPs to future work, and focus here on clarifying the experimental situation.

The outline of the paper is as follows.  In Section~\ref{sec:form} we review the formalism of direct detection of dark matter.  We review the characteristics of the relevant 
experiments, discuss the physics and implications of the channeling effect in DAMA, and discuss our analysis method.  We then apply these techniques to derive the allowed 
parameter space consistent with all experimental measurements for both elastically and inelastically scattered WIMPs in Section~\ref{numresults}.  Finally, we conclude and discuss future directions.

\section{Rates for direct detection of dark matter}
\label{sec:form}

\subsection{Direct detection formalism}

We first review the relevant features of direct detection of dark matter by WIMP-nucleus scattering.  The differential rate per unit detector mass in nuclear recoil energy is given by
\be
\frac{dR}{dE_R} = N_T \frac{\rho_{DM}}{m_{DM}} \int_{|\vec{v}|>v_{min}} d^3v\, vf(\vec{v},\vec{v_e}) \frac{d\sigma}{d E_R}.
\label{rate1}
\ee
Here, $N_T$ is the number of target nuclei per unit mass, $m_{DM}$ is the dark matter particle mass, and $\rho_{DM}= .3\, {\rm GeV/cm^3}$ is the local dark matter halo density.  $\vec{v}$ is the dark matter velocity in the frame of the Earth, $\vec{v_e}$ is the velocity of the Earth with resepect 
to the galactic halo, and $f(\vec{v},\vec{v_e})$ is the distribution function of dark matter particle velocities.  We take $f(\vec{v},\vec{v_e})$ to be 
a standard Maxwell-Boltzmann distribution:
\be
f(\vec{v},\vec{v_e}) = \frac{1}{(\pi v_0^2)^{3/2}} {\rm e}^{-(\vec{v}+\vec{v_e})^2/v_0^2}.
\ee
The Earth's speed relative to the galactic halo is $v_e=v_{\odot}+v_{orb}{\rm cos}\,\gamma\, {\rm cos}[\omega(t-t_0)]$ with $v_{\odot}=v_0+12\,{\rm km/s}$, 
$v_{orb}=30 {\rm km/s}$, ${\rm cos}\,\gamma=0.51$, $t_0={\rm June \, 2nd}$, and $\omega=2\pi/{\rm year}$.  We set the most probable dark matter 
speed in the galactic frame to $v_0={\rm 220 \,km/s}$ for most of our analysis, but study the effect of allowing it to vary within its 90\% C.L. range 
${\rm 170\, km/s} \leq v_0 \leq {\rm 270\,km/s}$~\cite{Kochanek:1995xv}.  The upper limit 
of the velocity integration of  Eq.~(\ref{rate1}) should be taken as the galactic escape velocity, ${\rm 490\,km/s} \leq v_{esc} \leq {\rm 730\,km/s}$ at 90\% C.L.~\cite{Kochanek:1995xv}.  
For simplicity we take the upper limit to infinity and {\it a postieri} restrict the minimum velocity to lie above $v_{esc}$.  We set $v_{esc} = {\rm 730\,
 km/s}$ for our study, but again study the effect of varying it within the 90\% C.L. range.  Neither uncertainty has a significant effect on our conclusions.  The minimum dark matter 
velocity $v_{min}$ depends on the recoil energy $E_R$ and the details of the direct detection process, and will be discussed later.

We assume a spin-independent cross section between dark matter particles and nuclei; a standard calculation~\cite{Jungman:1995df} leads to 
\be
\frac{d\sigma}{d E_R} = \frac{m_N}{2 v^2} \frac{\sigma_p}{\mu_n^2} \frac{\left[f_p Z+f_n (A-Z)\right]^2}{f_n^2} F^2(E_R).
\label{cross1}
\ee
$\mu_n$ is the reduced mass of the dark matter particle and nucleon (proton or neutron), $\sigma_p$ is the scattering cross section of the dark matter 
particle with nucleons, and $f_{n,p}$ are the coupling strengths of the dark matter particle to neutrons and protons respectively.  $f_{n,p}$ are 
calculated from a coherent sum over the couplings to the quark model constituents of the nucleon.  $Z$ and $A$ are the proton 
and atomic numbers of the nucleus, while $F(E_R)$ is a nuclear form factor which corrects for the above formula being strictly correct 
only as the momentum transfer $q^2=2m_N E_R \to 0$.  We use a standard Helm form factor~\cite{helm}.  

Inserting the dark matter-nucleus scattering cross section of Eq.~(\ref{cross1}) into the differential rate in Eq.~(\ref{rate1}) and performing the integration, 
we obtain the differential rate
\begin{eqnarray}
\frac{d R}{d E_R} &=& \frac{N_T m_N\rho_{DM} }{4 v_0 m_{DM}} \frac{\sigma_p}{\mu_n^2} \frac{[f_p Z + f_n(A-Z)^2]}{f_n^2}F^2(E_R) 
	\left\{\frac{{\rm erf}(x_{min}+\eta)-{\rm erf}(x_{min}-\eta)}{\eta}\right\} \nonumber \\ & & \times \Theta\left(v_{esc}-v_{min}(E_R)\right).
\label{rate2}
\end{eqnarray}
We have introduced the parameters $x_{min}=v_{min}/v_0$ and $\eta = v_e/v_0$.  The $\Theta$ function accounts for our treatment of the maximum 
allowed dark matter velocity, as discussed above.  We will use the formula in Eq.~(\ref{rate2}) to interpret the results of various direct detection searches 
for dark matter particles.  We follow the standard convention and set $f_p = f_n$; results for other exchange mechanisms, such as $Z$-boson 
exchange, can be derived by scaling the cross section $\sigma_p$ appropriately.  The total number of nuclear recoil events in a recoil energy range between $E_r^1$ and $E_r^2$ is 
\be
N = \sum_i \int_{E_r^1}^{E_r^2} \frac{d R_i}{d E_r} \frac{{\cal E}_i}{N_T M_i} dE_r
\label{rate3}
\ee
where the sum is over each nuclear species $i$ in the detector.  ${\cal E}_i$ is the effective exposure of species $i$ expressed in kg-days.   We can 
parametrize ${\cal E}_i = M_i t \epsilon$, where $t$ is the time of exposure, $M_i$ is the target mass of species $i$, and $\epsilon$ is a detection efficiency.

\subsection{Experimental constraints}

We discuss the direct detection signal obtained by DAMA and the constraints imposed by other null experiments.  The DAMA collaboration has recently reported an $8.2\sigma$ 
signal for an annual modulation signature of dark matter scattering~\cite{Bernabei:2008yi}.  This signal has been reported for several years, and the 
significance of the effect has been constantly increasing.  It is convenient to parametrize the DAMA detection rate as
\be
R_i = R_i^0 + S_i^1 {\rm \cos} [\omega(t - t_0)],
\ee
where $t_0={\rm June \, 2nd}= 152.5 \mbox{ days}$ is the time when the Earth is moving with its maximum speed with respect to the galactic halo and $T = 2 \pi/\omega = 1 \,{ yr}$.    The subscript $i$ denotes different energy bins.  The values measured by the collaboration in the $2-6\,{\rm keVee}$ bin are $t_0 = 144 \pm 8 \mbox{ days}$ and $T = 0.998 \pm 0.003 \mbox{ yr}$.  The keVee unit is keV-electron-equivalent, which accounts for quenching of unchanneled events; we describe this effect in detail in the next subsection.  For the moment, we only comment that $\mbox{keVee}=\mbox{keV}$ for channeled events, and $\mbox{keVee}=q \,\mbox{keV}$, where $q<1$, for unchanneled events.  The constant term $R_i^0$ is composed of both a signal piece coming from dark matter initiated processes, and a background piece arising from other sources of nuclear recoil: $R_i^0 = b_i^0+S_i^0$.    The expressions for $S_i^0$ and $S_i^1$ are obtained by integrating Eq.~(\ref{rate2}) over a given range of $E_R$.  More precisely, expanding the interaction rate in a Taylor series around $v_e = v_\odot$ for an energy interval between $E_i$ and $E_i + \Delta E_i$, we have
\be
R_i^0 = \frac{1}{\Delta E} \int_{E_i}^{E_i+ \Delta E_i} \left( \frac{dR}{dE_r} \right) d E_r
\ee
and
\be
R_i^1 = \frac{\Delta v_e}{\Delta E} \int_{E_i}^{E_i+ \Delta E_i} \frac{\partial}{\partial v_e}\left( \frac{dR}{dE_r} \right) d E_r,
\ee
where the differential rates are to be evaluated at $v_e = v_\odot$ and $\Delta v_e=v_{orb}{\rm cos}\,\gamma$.

The DAMA collaboration has not been able to find another effect besides dark matter scattering that could contribute 
to $S_i^1$.  The modulation amplitudes reported by DAMA in each energy bin are given in Table~\ref{DAMAamps}.  Non-zero 
modulation is clearly observed in the low energy bins, while it is absent in the $6-14$ keVee bin. 

\begin{table}[htbp]
\centering
\begin{tabular}{|c|c|} \hline
Energy & $S_i^1$ (cpd/kg/keVee) \\ \hline\hline
$2-4$ keVee & $0.0223 \pm 0.0027$ \\ \hline
$2-5$ keVee & $0.0178 \pm 0.0020$ \\ \hline
$2-6$ keVee & $0.0131 \pm 0.0016$ \\ \hline
$6-14$ keVee & $ 0.0009 \pm 0.0011$ \\ \hline\hline

\end{tabular}
\caption{\label{DAMAamps} Modulation amplitudes in units of counts-per-day/kilograms/keVee as reported by DAMA~\cite{Bernabei:2008yi}}
\end{table}

Any dark matter model allowed by the DAMA results must be consistent with other null experiments.  We consider constraints imposed by the CDMS, 
CRESST,  and XENON
collaborations.  We summarize the salient features of each experiment, including the target nucleus, energy threshold, and exposure, 
in Table~\ref{ExpFacts}.  Considering the constraints from a variety of experiments is vital, as each is relevant for different regions of dark matter 
parameter space.  We will see this explicitly in our numerical results of Section~\ref{numresults}.
The most important qualitative observation to make about the experiments is the difference in the energy thresholds.  The recoil energy is related to the mass of the incoming dark matter particle through the relation
\be
E_R = \frac{2 v^2 \mu_{nuc}^2}{m_N},
\label{thresh}
\ee
 where $\mu_{nuc}$ the reduced mass of the WIMP-{\em nucleus} system and $v$ their relative velocity.  It follows that for low mass WIMPs, the recoil energy of the nucleus is not high enough to be above threshold for XENON and CDMS, while it is for DAMA.    The experiments which do have thresholds competitive with DAMA, in particular the CRESST-I 
 experiment, have much lower exposures.  While they constrain the low dark matter mass region, their limits on the cross-section $\sigma_p$ are not strong enough  to constrain much of the DAMA parameter space.  We show in Section~\ref{numresults} that the CDMS-II Si run is also sensitive to lower dark mater masses because of the lower target mass of silicon and the lower threshold compared to the germanium data.
 
\begin{table}[tbhp]
\centering
\begin{tabular}{|c|c|c|c|c|c|} \hline
Experiment & Target & Exposure (kg-d) & Threshold  & Ref \\ \hline\hline
CDMS-SUF & Ge & 65.8 & 5 \mbox{ keV} & \cite{CDMSI} $$ \\ 
&                 Si   & 6.58 & 5 \mbox{ keV} & \\ \hline
CDMS-II & Ge & 121.3 & 10 \mbox{ keV} & \cite{CDMSII}\\ 
& Si & 12.1 & 7 \mbox{ keV} & \cite{CDMSIISi} \\ \hline
XENON10 & Xe & 131 & 4.5 \mbox{ keV} &  \cite{XENON}\\ \hline
CRESST-I & Al$_2$O$_3$ & 1.51 &  $0.6 \mbox{ keV}$ &  \cite{CRESST} \\ \hline
 \hline

\end{tabular}
\caption{\label{ExpFacts} Relevant features of the null experiments used in our analysis.}
\end{table}

\subsection{The physics of quenching and the DAMA signal}

Since it has important implications for dark matter searches, we discuss in more detail the physics of the channeling and quenching effects which allow the compatibility of the DAMA signal with other null experiments.   The relevance of this effect for the current 
analysis is that DAMA is sensitive to nuclear recoil energies down to 2 keV, below the threshold of both XENON and CDMS.    

The light yield of scintillators is different depending on whether the incident particles interact electromagnetically or via the nuclear force.  The bookkeeping for this is performed 
by introducing an electron equivalent recoil energy for a given nuclear recoil energy; for a nuclear recoil of 1 keV, the equivalent electron energy (in keVee) is 
$q_x \times 1 \,{\rm keV}$.  $q_x$ is the quenching factor for the nuclear material composing the scintillator; for the materials composing the DAMA detectors, $q_{Na} \approx 0.3$ and 
$q_I \approx 0.09$.  It is simple to understand why $q<1$ and the electron equivalent energy is less than that of the nuclear recoil energy.  A nucleus 
hitting scintillator material will lose energy both by electromagnetic and nuclear interactions, while an incoming electron will lose energy only via 
electromagnetic interactions. It is primarily the production of radiation in electromagnetic interactions that produces the light yield in the scintillating 
material, yielding $q<1$.

For crystal scintillators such as those used by DAMA, a portion of the events will be ``channeled," effectively changing the quenching factor to 
$q \approx 1$.  This occurs when incident particles that do possess nuclear interactions interact only electromagnetically with the scintillator material.  This 
can occur for certain energies and incidence angles of the incoming particle.  The importance of this for the DAMA experiment was first 
discussed in Ref.~\cite{Drobyshevski:2007zj}, and a detailed analysis of its effect was performed by the DAMA collaboration in 
Ref.~\cite{Bernabei:2007hw}; we refer the reader to these references for a more detailed discussion.  An estimate of the fraction of channeled events based on 
simulation is given in Fig. 4 of Ref.~\cite{Bernabei:2007hw}.  We use the following simple parametrization  in our analysis~\cite{Foot:2008nw}:
\be
f_{Na} \simeq \frac{1}{1+1.14 E_R(\mbox{keV})} , \mbox{        } f_{I} \simeq \frac{1}{1+0.75 E_R(\mbox{keV})} .
\ee

The other experiments considered in our study do not report this distinction between quenched and channeled events.  
CDMS collects all recoil energy from ionization and phonons (heat).
Channeling does not occur in liquid noble gases~\cite{Bernabei:2007hw}, such as used in 
XENON.  CRESST sets $q=1$ in their analysis.  
The XENON collaboration has incorporated the appropriate quenching 
factor into their published results.

\subsection{Analysis procedure}

We now discuss our analysis procedure for determining whether the DAMA results are consistent with the constraints of other experiments
for both elastic and inelastic dark matter scattering.  For experiments that report no events above background, we demand that dark matter 
initiated scattering produce less than 2.3 events throughout the entire range analyzed, as appropriate for establishing a 90\% C.L. upper limit.  For experiments that report 
events but ascribe them to background processes, we adopt a conservative approach and include the events when setting limits.  We determine the 90\% C.L. limit 
on the dark matter cross section in the presence of this background using a simplified version of the optimum interval method~\cite{Yellin:2002xd}.  We divide the data into energy bins using the published results.   We study all contiguous combinations of bins, and demand that the dark matter candidate produce no more than the 90\% C.L. upper limit allowed events as dictated by Poisson statistics in each such interval.  The most stringent limit in $\sigma_p$ for a given $m_{DM}$ among the studied intervals is then used.  This procedure allows effective use of the different kinematics of each dark matter particle when setting constraints, as the appropriate energy bins 
which constrain a given dark matter candidate depend strongly on the mass and whether elastic or inelastic scattering is considered.  Our technique reproduces 
reasonably well the published constraints in the elastic scattering limit, as obtained from either the experimental papers or {\small\sf DMTools}~\cite{dmt}.

For completeness, we now describe in detail our treatment of each experimental data set.

\begin{itemize}

\item{\underline{DAMA}:} 
For the DAMA signal, we perform a $\chi^2$ fit to the $2-6$ and $6-14$ keVee bins given in Table~\ref{DAMAamps} for a given $m_{DM}$ and 
$\sigma_p$:
\be
\chi^2 = \sum_i \left(\frac{R_i^1 - R_i^{1,exp}}{\sigma_i} \right)^2,
\ee
where $R_i^{1,exp}$ is the value of $R_i^1$ measured by DAMA in the $i$th energy bin, $R_i^1$ is the rate computed at a given $(m_{DM},\sigma_p)$ point, and $\sigma_i$ is the experimental error on $R_i^{1,exp}$.  The minimum $\chi^2$ is found at a given $m_{DM}$ by scanning over $\sigma_p$.  Once the minimum $\chi^2$ is found, if $\chi^2_{min}  < 2$, a 90\% C.L. region is determined by accepting values of $\sigma_p$ with $\chi^2-\chi^2_{min}<2.71$.  We also determine a $3\sigma$ allowed range for $\sigma_p$ by accepting cross sections leading to $\chi^2-\chi^2_{min}<9$.

\item{\underline{CDMS-II}:}
The latest results from the five-tower germanium run of CDMS indicate no events in a nuclear recoil energy range of $10-100 \,{\rm keV}$~\cite{CDMSII}.  
The efficiency is fairly flat, and it has already been included in the effective exposure given in Table~\ref{ExpFacts}.  
We demand that a dark matter candidate produce $<2.3$ events to establish a 90\% C.L. upper limit on $\sigma_p$ for a given $m_{DM}$.  We also include the constraint from the silicon run of Ref.~\cite{CDMSIISi}.  The energy window for this result is $7-100\,{\rm keV}$; the lighter mass of silicon 
and the lower energy threshold leads to important constraints for light dark matter.  We again demand that dark matter interactions 
produce $<2.3$ events.

\item{\underline{CDMS-SUF}:}
We include constraints from the CDMS run reported in Ref.~\cite{CDMSI}.  This early run has  implications for light dark matter because the analysis threshold was set at 5 keV.  The efficiency is not flat as a function of energy.  We use the parametrized form given in Ref.~\cite{Gondolo:2005hh}:
\begin{equation}
\epsilon = \left\{ \begin{array}{cc} 7.6\% & E<10\,{\rm keV} \\
				22.8\% & 10\,{\rm keV} < E < 20\,{\rm keV}\\
				38\% & E>20\,{\rm keV}. \end{array} \right.
				\end{equation}
This function must be included in the integration of Eq.~(\ref{rate3}).  The experiment does observe events in the expected dark matter signal range, 
although these are consistent with background expectations.  We determine the 90\% C.L. limit on the dark matter scattering rate in the presence 
of this background using a simplified version of the optimum interval method discussed above.  We divide the data into the 
four energy bins $E<10\,{\rm keV}$, $10\,{\rm keV} < E < 20\,{\rm keV}$, $20\,{\rm keV} < E < 55\,{\rm keV}$, and $E>55\,{\rm keV}$.  We obtain the background 
events in each bin from 
Ref.~\cite{CDMSI}.

\item{\underline{XENON}:}
We include the first results from the XENON dark matter search reported in Ref.~\cite{XENON}.  We again apply the optimum interval method 
using the efficiencies and energy bins given 
there, and the reported background events. 
All observed events 
are attributed to background processes by the collaboration, but we follow their analysis in including the observed events when computing a limit.

\item{\underline{CRESST}:}
The initial run of the CRESST experiment utilized a sapphire target with a low threshold of 0.6 keV~\cite{CRESST}.  We use the observed
spectrum without background subtraction given in 
that reference.  The collaboration attributes all observed events to background initiated processes, due to a study of coincident counts in 
multiple detectors.  We apply the optimum interval method with a restriction that the interval size be greater than 1.2 keV, which is 
twice the experimental energy resolution.  We use a 100\% efficiency for detection of events, as reported by the collaboration.

\item{\underline{TEXONO}:}
The TEXONO collaboration has recently reported results using a germanium target with a 200 eVee nuclear recoil energy threshold~\cite{TEXONO}.  The limits claimed in this paper (but questioned in~\cite{Avignone:2008xc}) do not yield significant improvement over CRESST-I.  We have checked that the $3-8\,{\rm GeV}$ WIMP mass window is still open with or without the inclusion of TEXONO constraints in the current analysis. Their projected sensitivity, however, cuts deeply into the DAMA parameter space.  We include the future projections in our analysis.  

\item{\underline{CoGeNT}:}
In the elastic scattering case we include the recent result of the CoGeNT collaboration~\cite{cogent}.  The current exclusion curve reported by the collaboration does not include the possible effect 
of channeling.  It is currently unclear by how much Ge-based ionization detectors such as used in CoGeNT are affected by channeling.  We caution the reader that the CoGeNT exclusion curves could drift on our plots when channeling in this type of detector is simulated.  Without channeling factored in, the measurement does not close the light WIMP window opened up by the 
channeling effect in the DAMA experiment, although future results from the collaboration should either confirm or severely constrain the DAMA parameter region.

\end{itemize}

\section{Implications for models of WIMP dark matter \label{numresults}}

Following the analysis procedure described in the previous section, we study whether parameter space exists in which spin-independent dark matter scattering 
off nuclei can simultaneously accommodate the DAMA modulation signal and satisfy the constraints from other experiments.  We begin by considering the case of elastic scattering.  We then examine whether inelastic scattering processes of the form $\chi_1 N \to \chi_2 N$, where $\chi_2$ can be either 
lighter or heavier than the dark matter particle $\chi_1$, can account for all experimental observations.

\subsection{Elastic scattering}

\begin{figure}[htbp]
\centering
\includegraphics[width=8.0cm,angle=90]{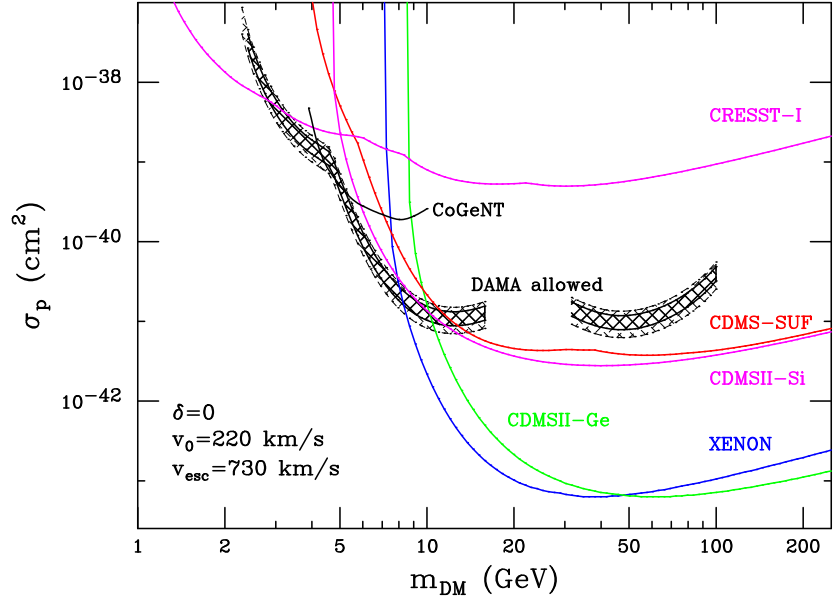}
\caption{\label{delta0}Allowed region in the $m_{DM}, \sigma_p$ plane consistent with the DAMA modulation signal at 90\% C.L. and $3\sigma$ (inner and outer hatched regions, 
	respectively).  Also shown are the experimental constraints arising from other null experiments.  The DAMA allowed region includes both channeled and quenched events.}
\end{figure}

We present in Fig.~(\ref{delta0}) the results of our parameter space scan for spin-independent dark matter scattering.  The inner and outer hatched regions respectively denote the 90\% C.L. and $3\sigma$ regions consistent with the DAMA modulation signal.  Regions above each colored line are excluded by the indicated experiments.  Although 
a portion of the region consistent with DAMA is excluded, a region with $m_{DM} \sim 3-8 \,{\rm GeV}$ is not excluded.  Two features of current 
experiments make this possible.  Substituting the 4.5 keV threshold for XENON and the 10 keV threshold for CDMS-II into Eq.~(\ref{thresh}) and setting 
$v=v_{esc}=730\,{\rm km/s}$, we see that XENON can only probe $m_{DM} > 7\,{\rm GeV}$, while CDMS-II germanium is sensitive to $m_{DM} > 8\,{\rm GeV}$.  Although these experiments have large exposures, the light dark matter region is inaccessible to them.  Further running of these experiments will not completely remove this window.  The 
experiments with low enough 
thresholds to be sensitive to the DAMA region, notably CRESST,
the 7 keV CDMS-II silicon run, and CoGeNT, do not have a large enough exposure, as seen from 
Table~\ref{ExpFacts}.  

\begin{figure}[htbp]
\centering
\includegraphics[width=8.0cm,angle=90]{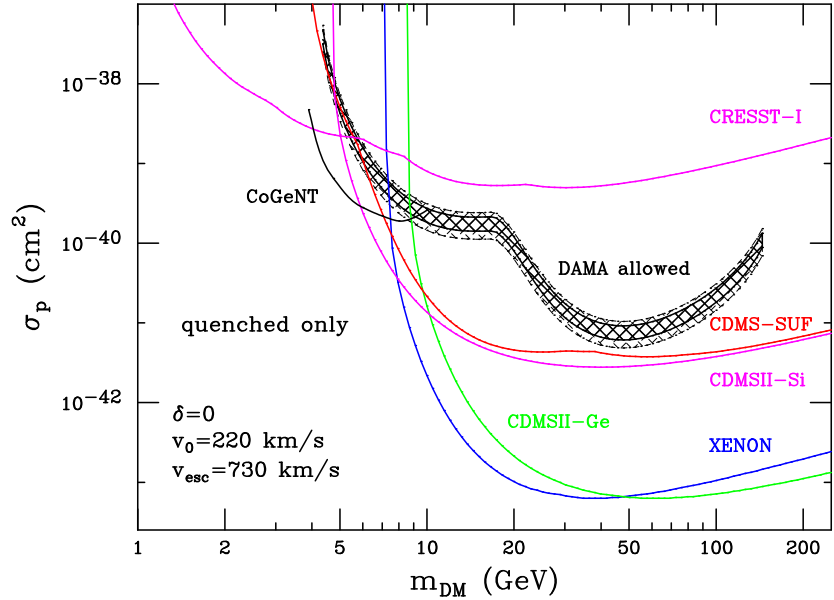}
\caption{\label{quench} Similar to Fig.~(\ref{delta0}), but if DAMA observed only quenched events.  The presence of un-quenched (channeled) events is necessary to reconcile DAMA with null experiments.}
\end{figure}

Several distinct structures are noticeable in the DAMA allowed region.  The low mass tail extending to large cross sections arises from channeled sodium events.  Both 
the low threshold and light nucleus cause these to appear first as $m_{DM}$ is increased.  The allowed region above $m_{DM} \approx 50\,{\rm GeV}$ arises from 
quenched iodine events.  The presence of channeling is crucial for reconciling the DAMA signal with other null experiments.  This is illustrated in Fig.~(\ref{quench}), 
where the effects of channeling have been removed.  No region in parameter space is consistent with all experiments when channeling is not considered.

The previous analysis of elastic scattering in Ref.~\cite{Gondolo:2005hh} utilized the light mass of sodium to render DAMA consistent with other experiments. 
At the time, all events in DAMA were considered to be quenched, leading to an approximately $2 \mbox{ keV}/q_I \approx 7 \mbox{ keV}$ energy threshold for scattering off sodium.  Although this 
threshold is similar to the thresholds of other experiments, sodium is lighter than other nuclei such as germanium, so that the net effect is sensitivity to lighter dark matter particles.  This led to the low-mass tail in Fig.~(\ref{quench}).  New limits from 
the 7 keV silicon run of CDMS-II have completely ruled out this region, although as shown in Ref.~\cite{Gondolo:2005hh} it is possible that fitting to the $2-4$ keV bin instead of the 
$2-6$ keV bin might open up a small amount of additional parameter space.  The 
channeling effect is required to open a significant low-mass elastic scattering window.

\begin{figure}[htbp]
\centering
\includegraphics[width=5.5cm,angle=90]{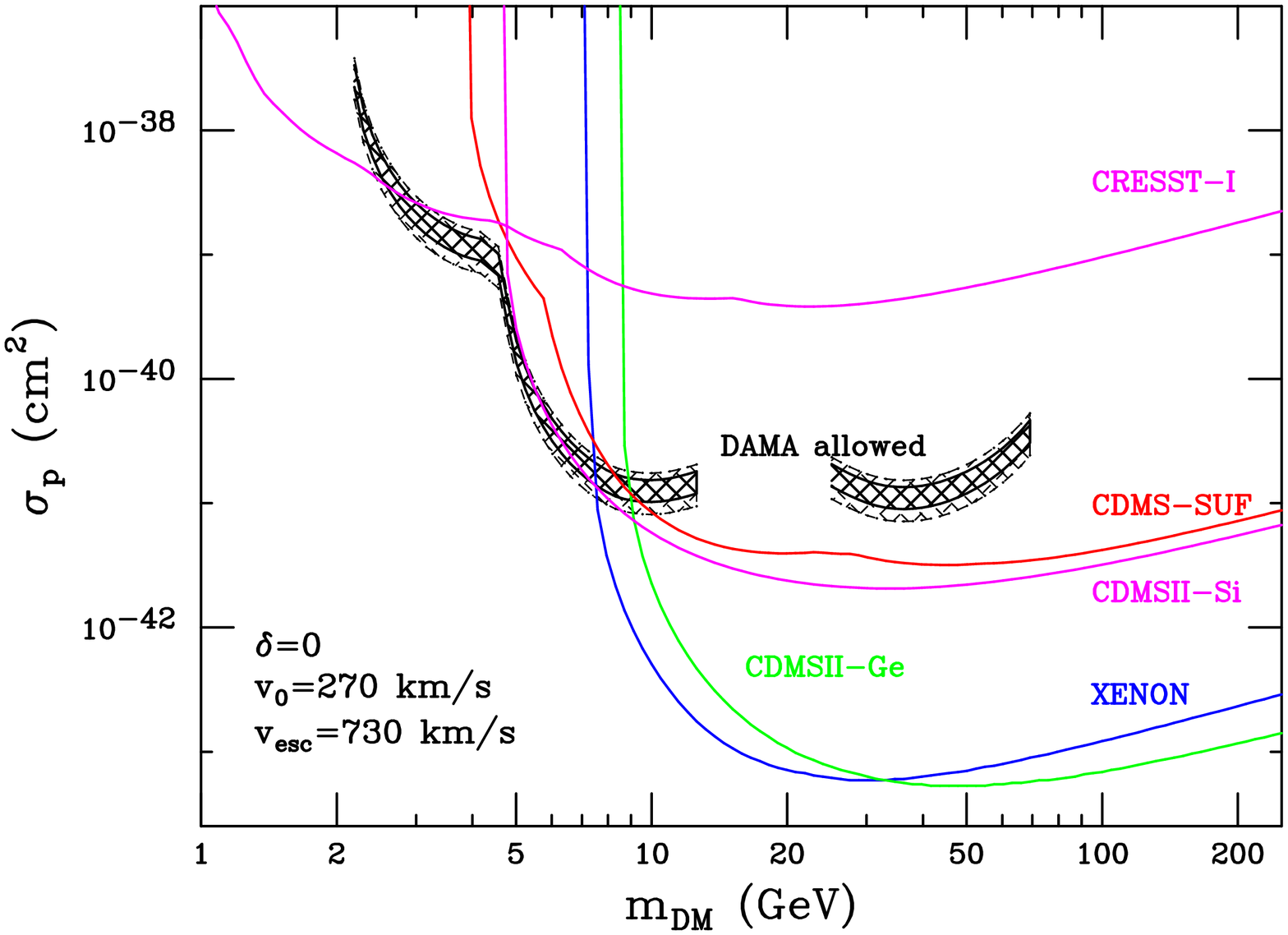}
\includegraphics[width=5.5cm,angle=90]{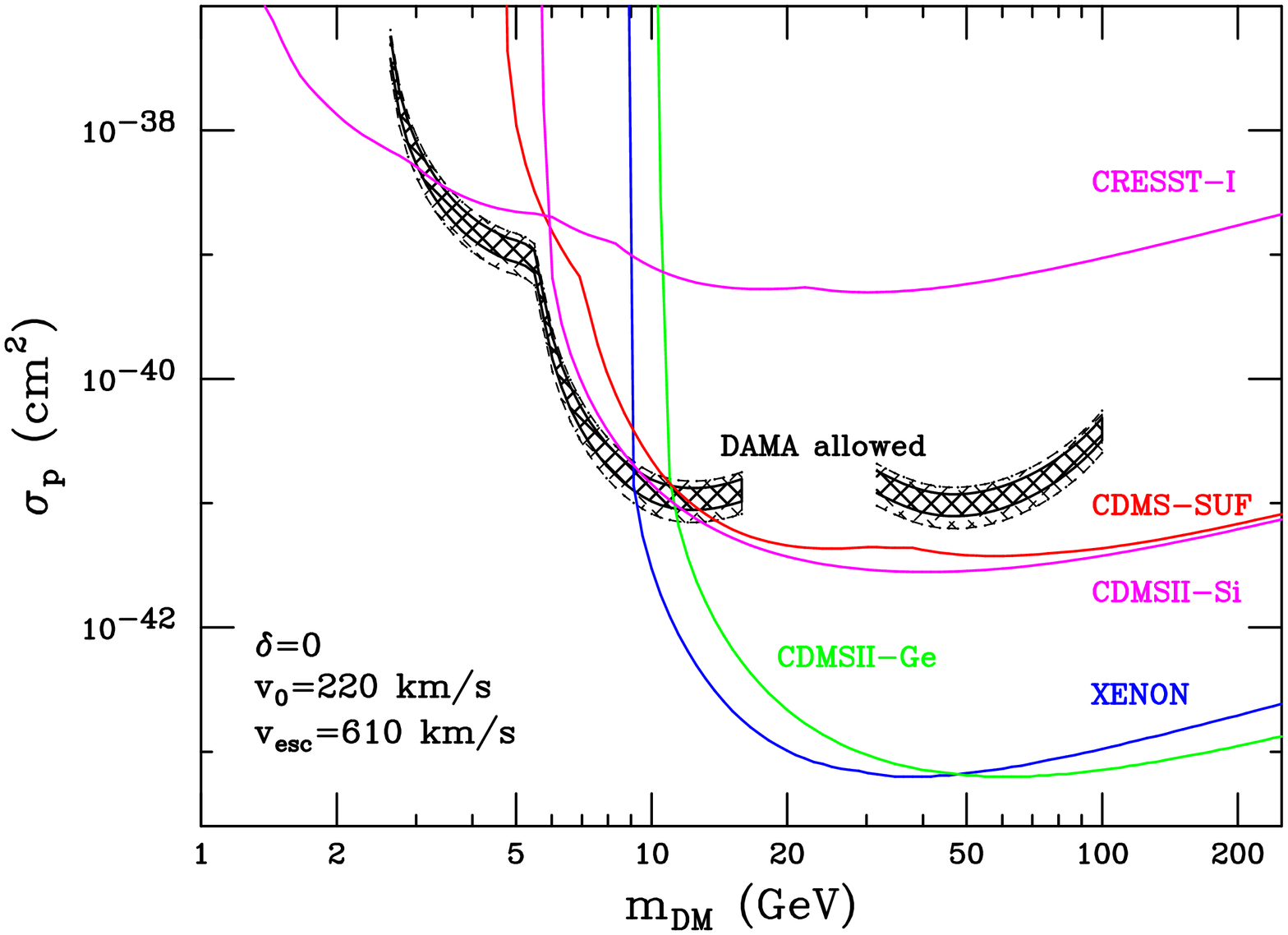}
\caption{\label{galpar} Region allowed by the DAMA modulation signal together with other experimental constriants obtained by changing to $v_0=270\,{\rm k/s}$ (left panel) and 
$v_{esc}=610\,{\rm km/s}$ (right panel).}
\end{figure}

To roughly study the effect of galactic uncertainties on these results, we also show in Fig. ~(\ref{galpar}) the effect of changing the most-probable speed $v_0$ and the galactic escape velocity.  In the left panel we change to $v_0=170\,{\rm km/s}$, while in the right panel we set $v_{esc}=610\,{\rm km/s}$.  Although both changes have quantitative 
effects, the picture described above is unchanged.  Previous analyses have shown that assuming a galactic velocity distribution beyond a simple Maxwell-Boltzmann form 
opens up a small region of additional allowed parameter space~\cite{Gondolo:2005hh}, and we expect the same would occur in our study, though we do not pursue it further here.

\begin{figure}[htbp]
\centering
\includegraphics[width=8.0cm,angle=90]{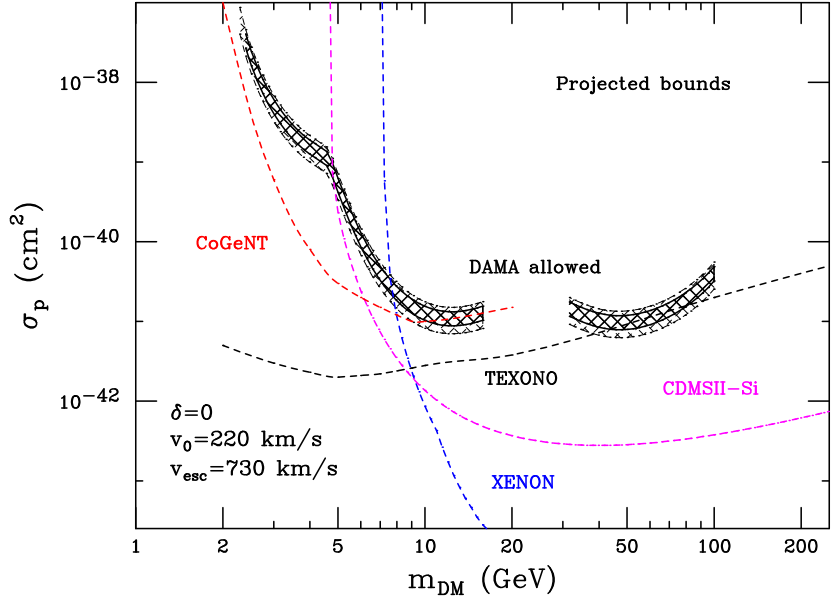}
\caption{\label{proj} Projected sensitivities of XENON and CDMSII-Si obtained by scaling the current exposures upwards by a factor of ten.  For XENON, both signal 
and observed background are scaled upwards,  For TEXONO, the projected sensitivity has been read off from~\cite{TEXONO}, while for CoGeNT the projected sensitivity is taken 
from~\cite{cogent}.}
\end{figure}

Further running of these experiments will continue to probe the elastic scattering parameter space.  To estimate whether future experiments can rule out the allowed region, we show in Fig.~(\ref{proj}) the projected sensitivities of XENON, CDMSII-Si, and TEXONO together with the DAMA allowed region.  These 
projections are obtained by simple scalings of the current exposure by a factor of ten.  For CDMS-II, the expected signal has been increased by ten, while for XENON the expected signal and observed background have both been scaled upward.  While the XENON energy threshold is currently too high to probe the DAMA allowed region, 
future CDMSII silicon data will test dark matter in the $5-8\,{\rm GeV}$ range.  For TEXONO, we have assumed an exposure of 1 kg-yr, the future 
goal of the collaboration~\cite{TEXONO}.  For CoGeNT, we have used the projected bound from Ref.~\cite{cogent} expected after an ongoing detector upgrade.  Achievement of these goals will allow both CoGeNT and TEXONO to study the entire low-mass WIMP window.

In summary, completely ordinary, elastically scattering WIMPs with masses in the $3-8\,{\rm GeV}$ range can produce the DAMA modulation signature and remain consistent with other experimental constraints.  Future data, particularly those from the silicon detectors of CDMSII, from TEXONO and from other low-threshold Germanium experiments such as used by the CoGeNT 
collaboration, are critically important to probe this parameter region.

\subsection{Inelastic scattering}

Inelastic scattering has also been considered as a mechanism for explaining the DAMA signal.  In models of this type, scattering processes $ \chi_1 N \to \chi_2 N$ are 
considered.  $N$ denotes the target nucleus and $\chi_1$ is the dark matter candidate.  The final-state particle $\chi_2$ can have a different mass than $\chi_1$.   We 
denote the mass splitting by $\delta \equiv m_2 - m_1$.  $\delta$ is positive for heavier final states, such as in the supersymmetric model of 
Refs.~\cite{Smith:2001hy,TuckerSmith:2004jv}.  Negative $\delta$ has been studied as a possible phenomenological explanation of the DAMA signal in 
Ref.~\cite{Bernabei:2008mv}.  

In the case of inelastic scattering, the maximum recoil energy can be shown by kinematics to be
\be
E_R^{max} = \frac{\mu_1}{m_{N}} \left\{ \mu_1 \beta^2-\delta +\sqrt{\mu_1^2 \beta^4-2 \mu_1\delta \beta^2}\right\}.
\label{recoilenergy}
\ee
Equivalently, for a given recoil energy, the dark matter must have a minimum velocity $v_{min}$ to scatter off a nucleus which is
\be
v_{min} = \sqrt{\frac{1}{2 m_N E_R}} \left( \frac{m_N E_R}{\mu_1}+\delta \right).
\ee
Here, $\mu_1=m_1 m_N/(m_1+m_N)$ is the reduced mass of the nucleus and $\chi_1$, and $\beta$ is the velocity of $\chi_1$ in units of $c$.  Inelastic dark matter with $\delta>0$ reconciled DAMA with the null experiments through two main effects \cite{Smith:2001hy,TuckerSmith:2004jv}.  First, for a sufficiently large mass splitting, only heavier targets can scatter inelastically for an experiment with a threshold $E_R^{th}$.  The 
restriction present in Eq.~(\ref{recoilenergy}) that the dark matter velocity $\beta^2>2\delta/\mu_1$ becomes increasingly stringent as $\mu_1$ is decreased.  Since iodine is relatively heavy in comparison to germanium, this could enhance the signal in DAMA relative to the germanium of CDMS.  This argument becomes much less effective with the presence of XENON constraints, since the mass of xenon is similar to iodine.  Second, positive $\delta$ changes the shape of the recoil energy spectrum, suppressing 
low energy recoils with respect to high energy ones.  Since before the discovery of channeling it was thought that the thresholds of DAMA were higher than those of the null experiments, DAMA favored inelastic dark matter.

\begin{figure}[htbp]
\centering
\includegraphics[width=5.5cm,angle=90]{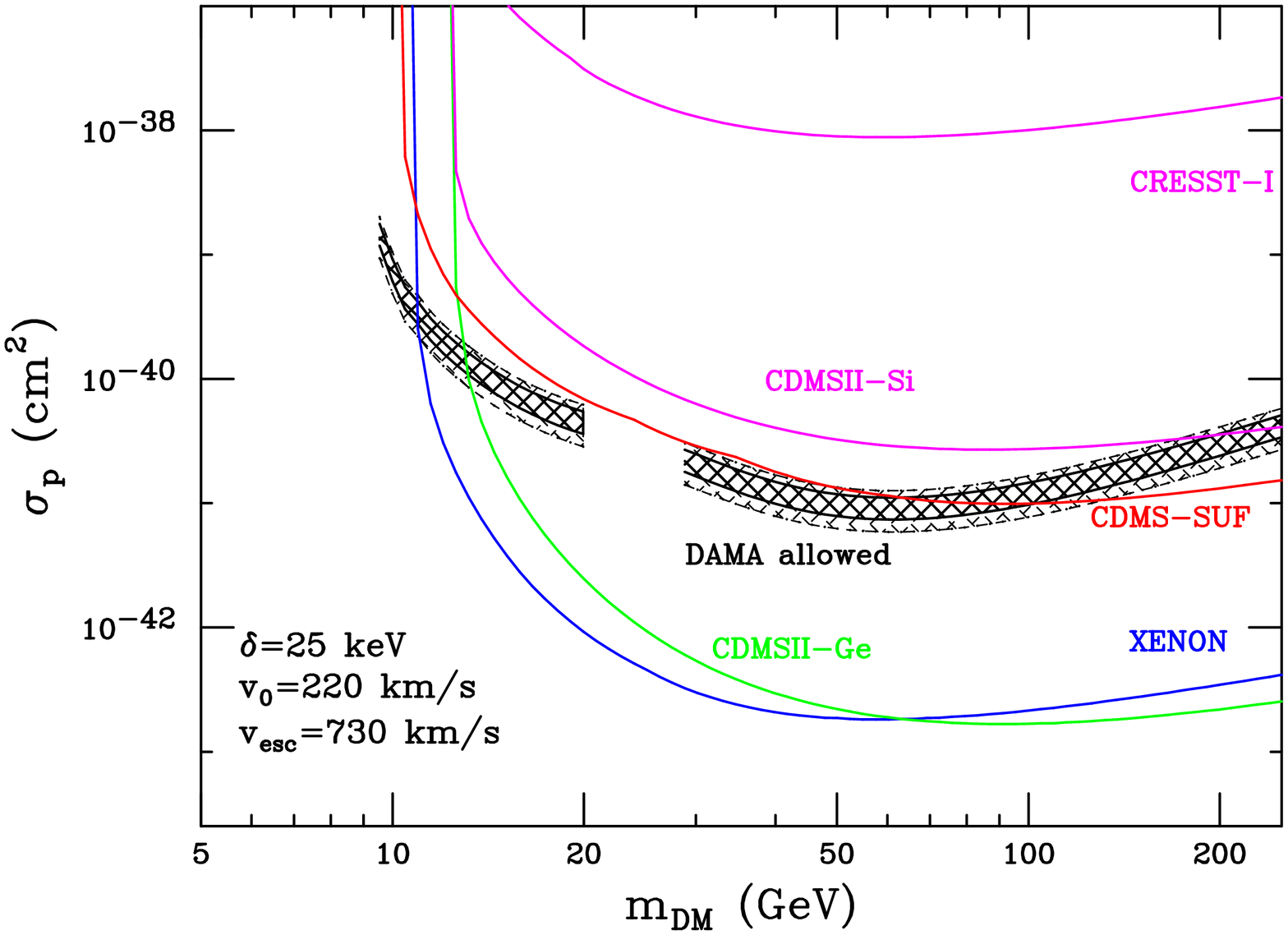}
\includegraphics[width=5.5cm,angle=90]{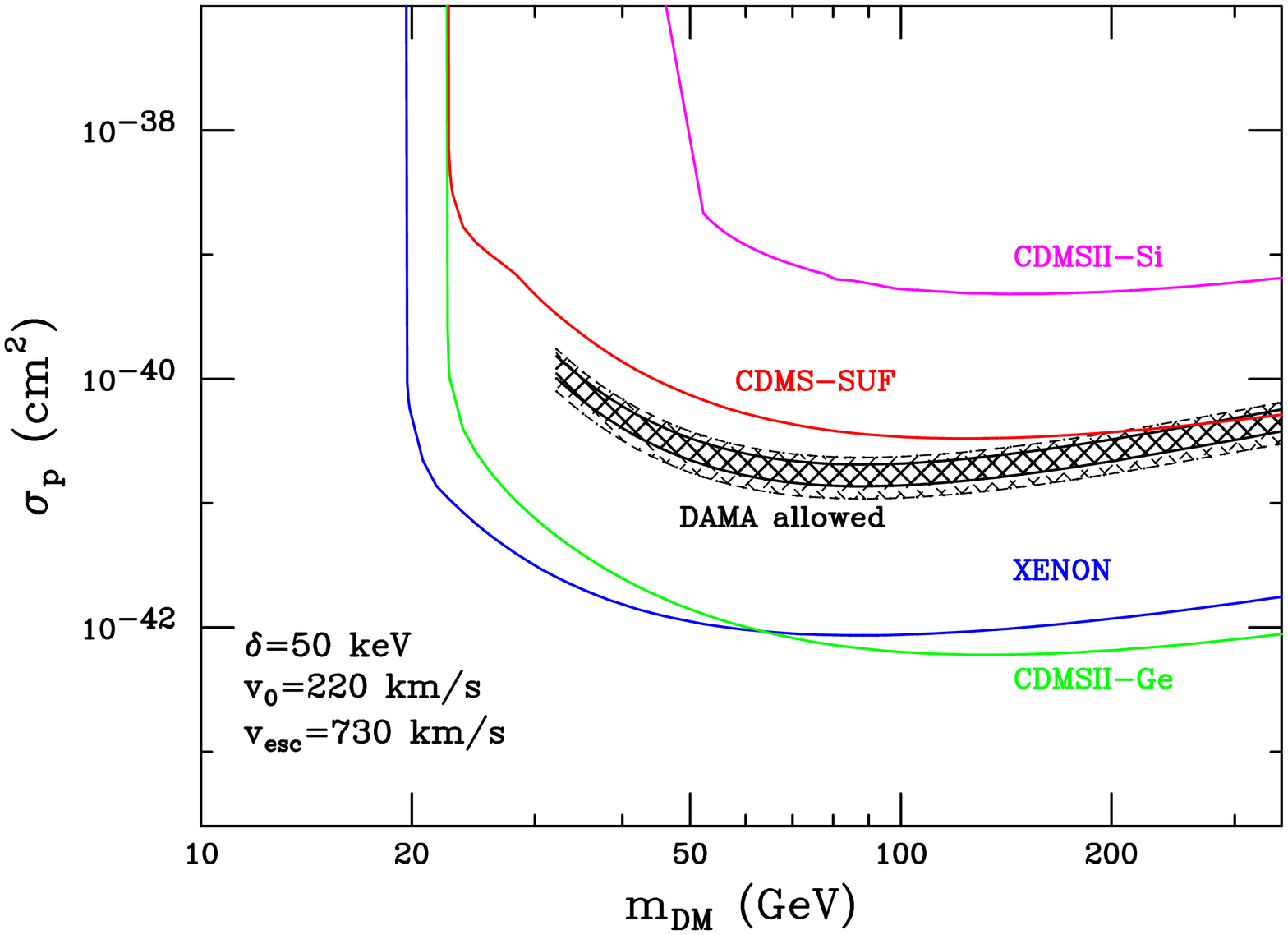}
\caption{\label{inelasticp} Region allowed by the DAMA modulation signal together with other experimental constriants in the $m_{DM},\sigma_p$ plane for $\delta=25,50\,{\rm keV}$.}
\end{figure}

However, we find that with the inclusion of channeled events and the new constraints from XENON and CDMS-II, inelastic dark matter becomes severely constrained.   We 
show in Fig.~(\ref{inelasticp}) the DAMA allowed parameter space together with other experimental constraints for the representative values $\delta=25,50\,{\rm keV}$.  The region of weak scale dark matter $m_{DM} \sim 100 \mbox{ GeV}$ and 100 keV mass splittings proposed in \cite{Smith:2001hy,TuckerSmith:2004jv} is closed.  The 
new experimental constraints from XENON and the germanium run of CDMS-II are too severe.  The small sliver of allowed parameter space present in the $\delta=25\,{\rm keV}$ case arises from channeled scatterings from iodine.  Although some regions for not-too-large $\delta$ are still consistent with all measurements, inelastic 
scattering with $\delta>0$ does not open up significant parameter space beyond elastic scattering.

There are two additional features to note in  Fig.~(\ref{inelasticp}).   First, relative to the $\delta = 0 $ case, positive $\delta$ shifts the region allowed by DAMA to larger masses.  This occurs because of the higher incident energy required to push the scattering over threshold.  Second, the constraints from experiments with light targets such as 
CRESST become much less significant.  The constraint on the initial dark matter velocity $\beta^2>2\delta/\mu_1$ discussed before indicates that 
for light nuclei, only the high velocity tail of the Maxwell-Boltzmann distribution can lead to dark matter scattering.

\begin{figure}[htbp]
\centering
\includegraphics[width=5.5cm,angle=90]{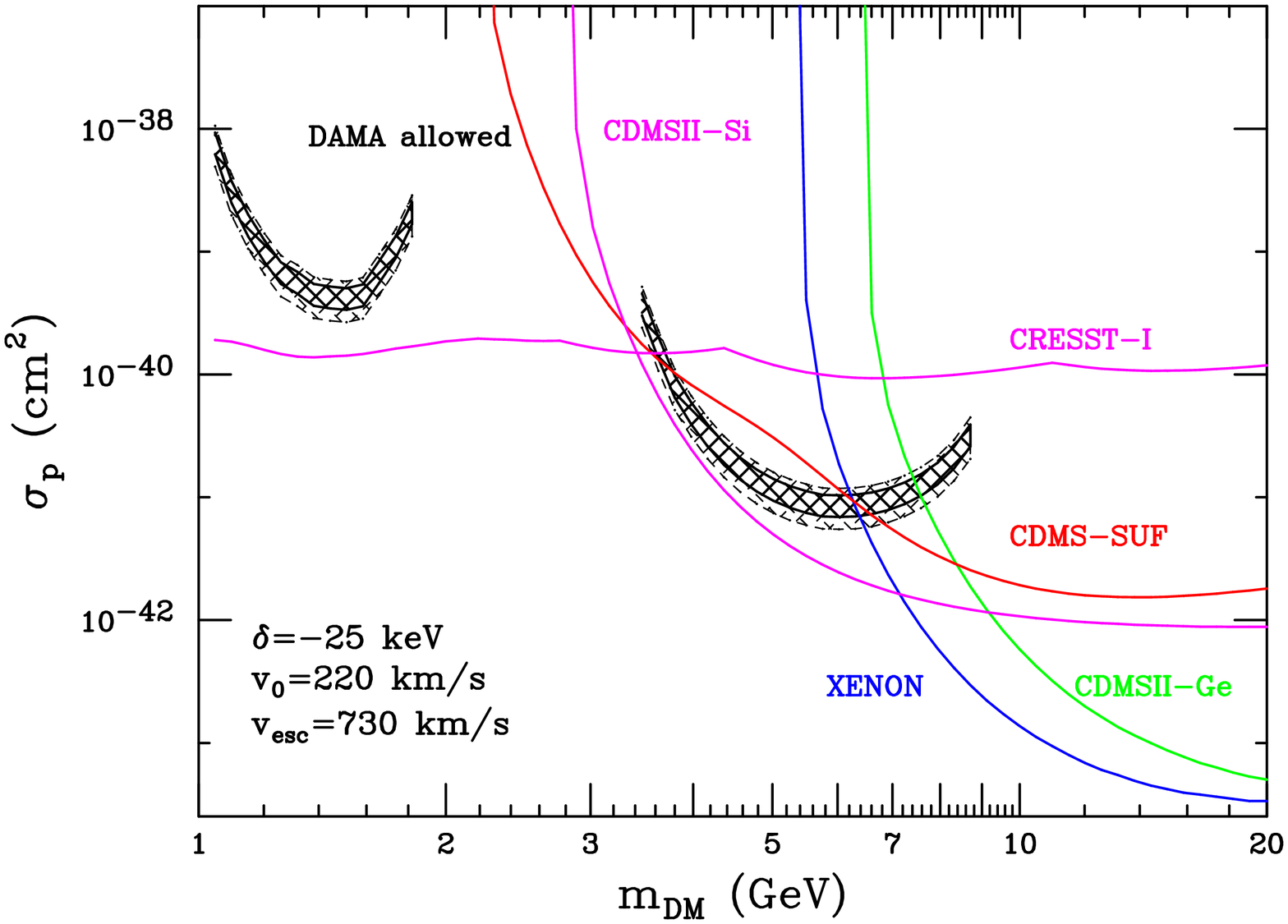}
\includegraphics[width=5.5cm,angle=90]{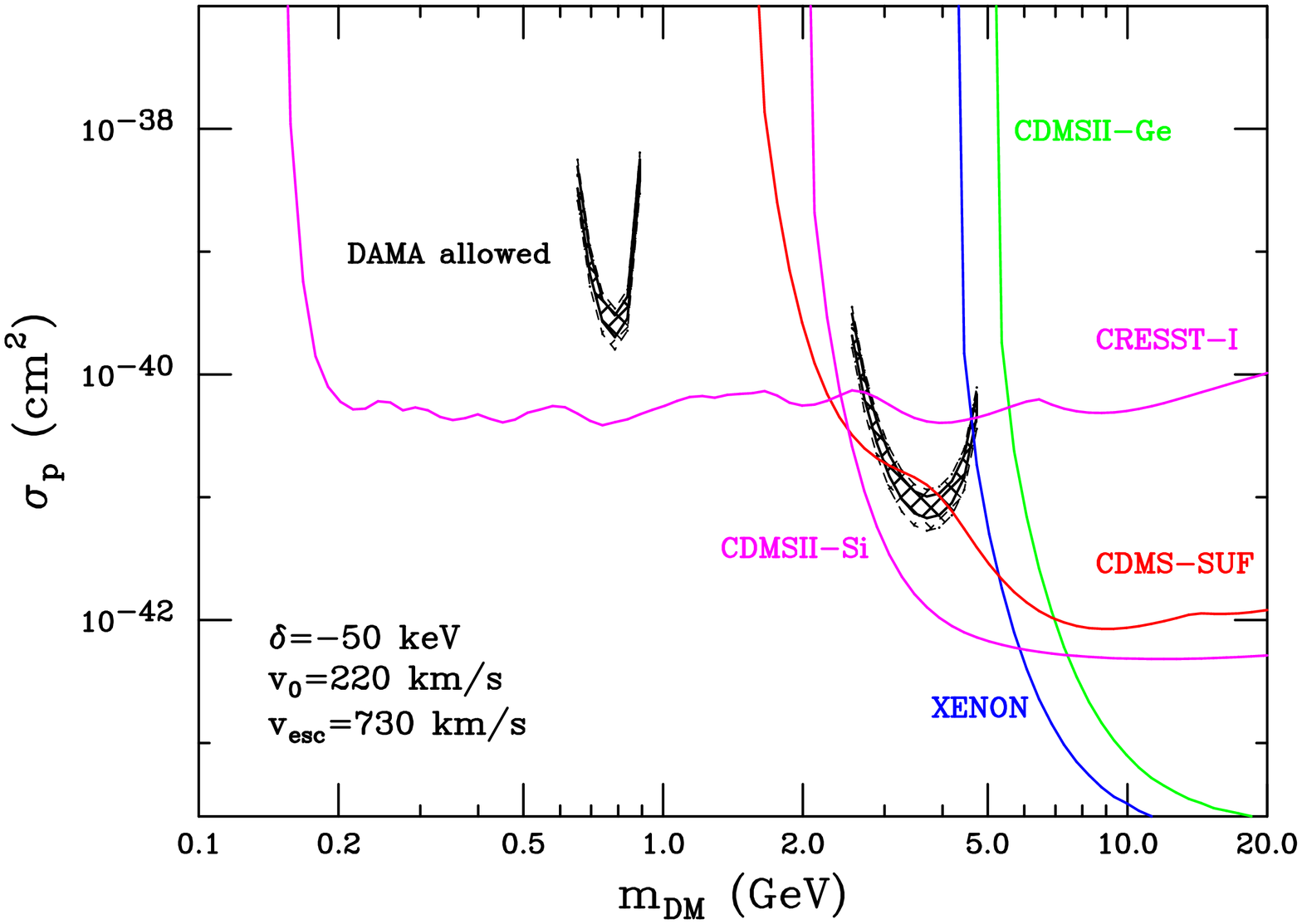}
\caption{\label{inelasticm} Region allowed by the DAMA modulation signal together with other experimental constriants in the $m_{DM},\sigma_p$ plane for $\delta=-25,-50\,{\rm keV}$.}
\end{figure}

We now study the region $\delta<0$.  We display in Fig.~(\ref{inelasticm}) the DAMA allowed parameter space together with other experimental constraints for the representative values $\delta=-25,-50\,{\rm keV}$.  The DAMA allowed region shifts to lower dark matter masses.  The lighter-mass horseshoe-shaped region is associated with channeled 
events from sodium, while the region for slightly higher masses comes from both quenched sodium and channeled iodine events.  The lack of an allowed region at 
masses higher than roughly 10 GeV occurs because the phase of the DAMA signal cannot be correctly obtained; a minimum in the dark matter scattering rate is predicted at 
$t_0=\textrm{June 2nd}$ rather than a maximum.  Only a small sliver of the higher-mass region is consistent with all measurements, with the light-mass parameter space arising from sodium events ruled out by CRESST.  The increasing severity of the these constraints as $\delta$ is descreased can be understood by studying the regions of the Maxwell-Boltzmann distribution contributing to dark matter scattering as a function of $\delta$.  Only a small portion of the Maxwell-Boltzmann 
distribution can initiate scattering for $\delta \geq 0$.  Lower velocities very quickly begin to lead to scattering as 
$\delta$ is decreased, strengthening the constraints for negative $\delta$.  The sharp turn-off of the CRESST
constraints at low dark matter masses 
visible in Fig.~(\ref{inelasticm}) follows from the fact that Eq.~(\ref{recoilenergy}) requires that $m_1 \gtrsim m_N E_R^{thresh}/(\sqrt{2m_N E_R^{thresh} \beta_{esc}^2}+|\delta|)$ for dark mater particles significantly lighter than the target nucleus.  We note that the wiggles visible in the exclusion curves arise from our use of 
the binned optimum interval method; as the kinematics of the dark matter scattering changes, a different energy interval is selected to provide the most stringent constraint.

\begin{figure}[htbp]
\centering
\includegraphics[width=8cm,angle=90]{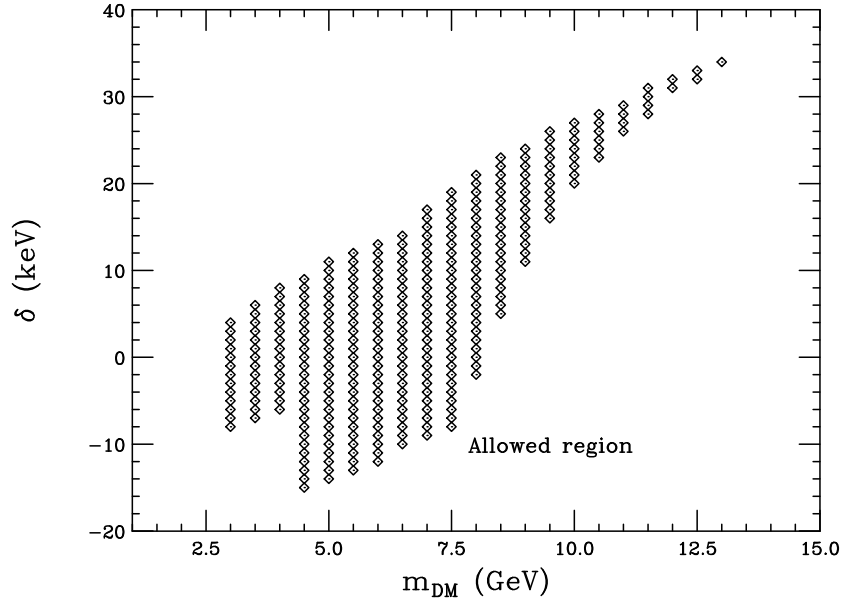}
\caption{\label{mdmdelta} Region of parameter space in the $m_{DM},\delta$ plane where the DAMA signal is consistent with the constraints from all null experiments.  The widest allowed range of dark matter masses occurs where $\delta = 0$.}
\end{figure}

Finally, we show in Fig.~(\ref{mdmdelta}) the region in the $m_{DM},\delta$ plane consistent with both the DAMA signal and all considered experimental constraints.  Dark matter 
masses in the region $3-13 \,{\rm GeV}$ and mass splittings in the range $-15 \,{\rm keV} \lesssim \delta \lesssim 30\, {\rm keV}$ are allowed.  We note, however, that the precise width of these intervals is quite sensitive to small deviations in the constraint curves and in the signal.  A combination of constraints from all null experiments is critical in obtaining this picture; XENON and CDMSII-Ge constraints are most important for positive $\delta$, while CRESST and CDMSII-Si
close off the negative $\delta$ window.  The largest range of allowed dark matter masses occurs for $\delta \approx 0$, indicating that inelastic scattering processes are not very helpful in reconciling the DAMA signal with other experimental constraints.

\section{Conclusions}

We have studied the consistency of the dark matter interpretation of the annual modulation signal observed by DAMA with the results of the null 
experiments CDMS, CRESST, 
and XENON.  Recent work has shown the presence of a channeling effect in the crystal scintillators utilized by DAMA which drastically changes the interpretation of the experimental results.  The presence of the channeling effect opens a window in dark matter parameter space between 3 and 8 GeV where the DAMA signal is consistent with all of the null experiments.  This consistency requires no exotic dark matter physics--a vanilla, elastically scattering dark matter candidate interacting through spin-independent channels is sufficient to explain both the signal and the null results from the other experiments.  We have also examined 
whether possible inelastic processes can accommodate all experimental results.  Inelastic scattering of dark matter particles to heavier final states renders
dark matter masses up to approximately 13 GeV consistent with all measurements.  However, the largest range of permissible dark matter masses occurs for elastic scattering 
candidates, indicating that inelastic processes do not open up significant regions of parameter space.  Future measurements from ultra-low energy germanium detectors and silicon results 
from CDMS are needed to explore this light-mass window.  

The light dark matter window suggested by the DAMA results motivates many new directions for model-building and phenomenology with low-mass WIMP 
candidates.  Although more model-dependent, the implications of light dark matter  for indirect detection and collider experiments should be explored.  It would be interesting to also consider whether spin-dependent scattering allows a larger range of dark matter masses to be consistent with DAMA and the 
null experiments.  

\section*{Acknowledgments}

We thank H. T. Wong for helpful correspondence regarding the TEXONO experiment, and J. Collar for comments and helpful communication regarding ultra-low energy germanium detectors.  The authors are supported by the DOE grant DE-FG02-95ER40896, Outstanding  Junior Investigator Award, 
by the University of Wisconsin Research Committee
with funds provided by the Wisconsin Alumni Research Foundation, and
by the Alfred P.~Sloan Foundation.

\medskip


\end{document}